\documentstyle[12pt]{article}
\begin{document}
\begin{center}
{\Large Does brane cosmology have realistic principles?}\\
\bigskip
{\bf D.H. Coule}\\
\bigskip
 School of Computer Sciences and Mathematics,\\ University of Portsmouth, Mercantile
House, Hampshire Terrace, Portsmouth PO1 2EG.\\
\bigskip
\begin{abstract}
The  maximal symmetry, or Perfect Cosmological Principle(PCP),
that prevents AdS type spaces from degenerating into
anti-inflationary collapse is argued to be unphysical. For
example, the simple requirement that brane-bulk models should be
the result of having evolved from even more energetic string
phenomena picks out a preferred time direction.
 We question whether quantum cosmological
reasoning can be applied in any meaningful way to obtain, what are
essentially, classical constructs . An alternative scheme is to
more readily accept the PCP and allow the branes to also become
eternal. A perpetually expanding and contracting brane model could
be driven by the presence of charged black holes in the AdS bulk,
that effectively violates the weak-energy condition as
singularities are approached. This can be contrasted with the
so-called Ekpyrotic universe which also closely accepts the PCP.
This being broken only by occasional collisions between branes,
that can then simulate a big bang cosmology.
\\


\end{abstract}
\end{center}
\newpage
{\bf Introduction}

 There is presently a lot of interest in brane theories and their
 cosmological
 properties.  Typically these models and those including   Randall and Sundrum
have branes existing in bulk spaces with  extra, possibly
non-compact, dimensions [1]. We will not  be too specific as to
which brane model we consider  since  many variations  have been
presented. Instead we address some common concerns to aspects that
seem generic in many  models, for a review and extensive list of
references see [2]. For this reason the concerns will be mostly
independent of the actual compactification scheme and orbifold
symmetry implored. We further only consider models that are
effectively five dimensional,with the Planck scales related by
\begin{equation}
M_4^2=M^3_5 V
\end{equation}
where $M_5$ and $M_4$ are the 5 and 4 dimensional Planck masses
and  $V$ the ``volume'' of the extra dimension. By having a
relatively large $V$ one can obtain a much smaller 5 dimensional
Planck scale, say TeV, compared to the usual 4 dimensional one
$\sim 10^{19}GeV$ [1,2].

 We also particularly have in mind models where positive and negative
cosmological constants are both present. A specific example is
when the positive tension of the brane is typically balanced with
a bulk negative cosmological constant. This results in static or
global Anti-de Sitter (AdS) space, or black hole variations
thereof, being the bulk space time.  This will be argued is
unjustified and counter to the requirements of fitting string
theory within a consistent cosmological model.

The five dimensional model does not live in isolation. It is the
result of earlier evolution from an even higher dimensional phase,
presumably starting with 11 dimensional M theory. The spacetime
should follow a dynamical evolution and not simply be imposed as
static, which implies it has always existed and remains unchanging
. The metric might become stabilized by new matter components
causing such behaviour. But the metric itself should not be
constrained initially in such a way. As the models are presently
constructed the  Perfect Cosmological Principle (PCP) is imposed
on the 5 dimensional bulk theory but only the  Cosmological
Principle (CP) is believed valid in our actual  4 dimensional
brane universe. Recall that while the CP takes our place in space
as not distinguishable, the PCP extends this also to time, see
eg.[3]. There is a double standard which is similar to an earlier
discussion regarding the distinction between a de Sitter solution,
which can also obey the PCP, and a physically inspired
inflationary one. For similar reasons that discounted pure de
Sitter we wish to discount Anti-de Sitter and replace it with the
slightly  less symmetric but now more realistic  form that only
satisfies the Cosmological Principle. This results in, and
probably fatally, a rapid collapsing solution, or what we can term
anti-inflationary behaviour. Note we avoid the use of the term
deflation, since this usually refers to the exponentially
collapsing stage of a closed de Sitter model.

{\bf De Sitter vs Inflation}

 In an earlier discussion Ellis and Rothman [4] pointed out that
 de Sitter space caused by a cosmological constant $\Lambda$ is
 unsatisfactory since it displays no preferred direction of time.
 This is due to the symmetries of the metric involving ``isometries
 transitive on space-time''[5] and constant-time surfaces are
 arbitrary. However, the presence of matter introduces a preferred
 time direction which breaks the full de
  Sitter $SO(4,1)$ symmetry in 4 dimensions, to those
 like  Friedmann-Robertson-Walker (FRW), which  only involves
 ``isometries transitive on space-like surfaces''. These isometry subgroups
  are of the form $SO(4)$, $SO(3,1)$ and E(3) for closed, open and
  flat spaces respectively. The quantum vacuum states of massless scalar
  fields also display this reduced symmetry [6].

 In inflationary models, we generally
 have a matter component that simulates a cosmological constant.  This is the case
 for a universe driven by a scalar potential $V(\phi)$, where the
 small kinetic term can still
  define a time like four velocity vector $U^{\mu}$ [7]. This vector in
 turn means that the full symmetry of pure de Sitter is broken and
 that the  solution actually corresponds to  exponential  expansion
 of the universe. Incidentally, the requirement  of the four velocity
 vector can place constraints on the spatial gradient terms of the
 scalar field [7].

 Regardless of a matter component that breaks the full symmetry of
 the de Sitter group, there is a second reason why the
 inflationary solution is to be preferred. We require the universe
 to later have a dynamical evolution described by the FRW metric
 with a global time parameter [7]. Introducing such a parameter, by fiat,  at the
 end of some static phase would simply re-introduce a causality problem that
 inflation is supposedly meant to overcome. By starting with an
 expanding inflationary solution this problem is prevented
 although at the expense of first choosing a specific choice of
 coordinates to represent this expansion. In other words a
  preferred time like direction should be initially chosen since
  one is anyway needed in the FRW universe, or any universe
  with evolution.

  In summary, the full de
 Sitter symmetry should be reduced because a) matter determines a
 time like direction and b) we require matching to a dynamical
 space time.

{\bf Anti-de Sitter vs Anti-inflation}

We can make a similar distinction between a pure AdS space and
that of a negative $\Lambda$ force that will accentuate
gravitational collapse: an anti-inflationary rapid collapse. The
AdS space is generally  taken as  static or global  when present
in string and supersymmetric theories [8]. The corresponding
isometry symmetry group is $SO(n,2)$ for $n+1$ dimensional AdS
space-see eg.[9]. It can be represented in the static form
\begin{equation}
ds^2= -\left( k+\frac{r^2}{l^2}\right )
dt^2+\left(k+\frac{r^2}{l^2}\right ) ^{-1} dr^2 + r^2 d\Sigma ^2_k
\end{equation}
where the negative cosmological constant is defined as $\Lambda=
-1/l^2$, with $l$  the curvature scale. For $k=$  $1,0,-1$,
$d\Sigma^2_{k}$ represents the unit sphere, plane and hyperbolic
space in $n-1$ dimensions respectively. Strictly speaking the full
AdS metric only corresponds to the $k=1$ case. For the Penrose
diagrams of odd dimensional AdS space-times see eg. [10], where
$k=1$ corresponds to $M=-1$ in their notation. There is a slight
complication, Brane models mostly take the $k=0$ form of the
metric, or the case $M=0$ in ref.[10]. This metric, in so-called
Poincare form, apart from not entirely covering the manifold has
problems when $r=0$, and is especially susceptible to
singularities if any matter component is present [10]. For this
reason Brane models typically exclude the $r=0$ regions of two AdS
spaces  and then ``paste'' the resulting manifold  together- see
eg.[11] for the relevant Penrose diagrams.

 For similar reasons why the dynamical inflationary
metric was justified we will reason that this AdS symmetry should
be reduced to give anti-inflation in actual string models.
Firstly, the bulk is never entirely empty since dilatons and
moduli fields also propagate there. Such matter fields can be used
to introduce a time-like direction into the model just as a scalar
field does in inflationary models. Secondly, the 5 dimensional
effective theory is the result of some earlier stage evolving from
even higher energy scales when the full M theory will presumably
be required. A  time evolution is already present before the AdS
stage becomes relevant. In the inflationary case the time
evolution was strictly required only after the inflationary stage;
but this, was argued, made its introduction necessary from the
start if inflation was to be useful. Breaking the static nature of
the metric reduces the isometry to $SO(n,1)$ so that the metric is
given by [8,9,12]
 \begin{equation}
 ds^2=-dt^2+cos^2(t) d\Sigma^2_n
 \end{equation}
 with $d\Sigma^2_n$ only now the metric on the unit hyperbolic space in n
 dimensions.
 There is a third reason why the reduced symmetry
 of only homogeneity and isotropy $SO(n,1)$  can be justified.
The anti-inflation spacetime is now globally hyperbolic, and the
metric covers the domain of dependence of the, say , $t=0$ initial
( now a Cauchy ) surface [8,9]. This is now consistent with the 5
dimensional phase evolving from some earlier regime and which
should be entirely determined by it. Unlike the full $k=1$ AdS
space there is no time-like surface at spatial infinity present
which prevents a suitable Cauchy surface. Information before could
enter the spacetime from the time-like or naked singularity
surface. Although, it might be possible to quantize such
non-globally hyperbolic metrics by controlling information
entering from such a surface [13], such boundary conditions are
difficult to justify and more severe than the usual horizon
problems of FRW metrics. One is setting up a condition on the
surface that will actually hold for all the time the model exists.
It is doubtful, that for example quantum cosmology using the
Wheeler-DeWitt equation would be sufficient. As presently
formulated such schemes generally assume a foliation of space or
an implicit time parameter that provides initial condition on a
suitable Cauchy surface - see eg.[14,15]. Singularities are
usually avoided or regulated in some way. Naked singularities are
further discounted  as being forbidden by the cosmic censor [16],
although this isn't fully understood directly from quantum
gravity. Naked singularities would however be difficult to
implement in quantum cosmology. In comparison, quantum cosmology
can be easily applied to the case of negative $\Lambda$ within an
open  FRW ansatz, as for example in ref.[17] . This is not very
helpful since the model is classically already prone to rapid
collapse and suffers from the same problems (eg. horizon and
flatness) as any non-inflationary FRW universe.

 Now, it might be
claimed that ``cutting and pasting'' the $k=0$ or $k= -1$ Ads
metrics together can still allow  a Cauchy surface cf.[18], and
there is little advantage in the FRW AdS metric in this regard.
Note this options would only occurs for odd dimensional
space-times, since we require the ``triangle'' Penrose diagram of
ref.[10]. However, this is still at the expense of non-causal
``cutting and pasting'' which is equally problematic. The
arguments above can equally be levelled against this ``process''
that conveniently ``knows how to remove the bad parts of the
metric'' and patch copies together if geodesic completeness [8] is
to be maintained. We will generally address our criticism to the
use of full AdS space where the lack of a Cauchy surface is
explicit, but bear in mind this can also be levelled against these
other schemes.

 In
summary, we can justify the use of the cosmological form of AdS
for similar reasons that a de Sitter phase is an over idealized
version of inflation; the presence of certain matter fields and
evolution from an earlier higher dimensional phase which breaks
the full AdS symmetry to that, at best,  like FRW. Although, these
coordinates do not cover the whole of AdS they are more physically
justified for the reasons given .  This distinction is  similar to
the difference between a true black hole formed by collapsing
matter and that of the ideal eternally existing static case. The
left hand side, or other ``universe'' , of the Penrose diagram is
discounted in the physical collapsing star case, although it is
present in the maximally extended Schwarzschild metric [8].

  There
is now  also the added advantage that a Cauchy surface is present
at the start of this cosmological AdS behaviour.  Losing the
static nature seems to be fatal as now gravity is simply being
made more attractive and rapid contraction can occur. Explaining,
the still highly symmetrical ansatz of homogeneity and isotropy
seems an added complication if inflation is not to be invoked.  We
will next consider some specific implications of using only the
cosmological metric form of AdS in brane cosmology and holography
ideas.

{\bf Brane cosmology}

We address our concerns to higher dimensional models where
positive and negative cosmological constants play a crucial role
in the formulism. We will make particular use of the approach of
ref. [19]. Although, this does not exclude all possible schemes it
does relate to the Randall Sundrum model where a ``see-saw''
mechanism relates the various cosmological constants. For example,
in a 5 dimensional theory one obtains  a relation [2,19]
\begin{equation}
\Lambda_4 =\Lambda_5+ \lambda^2
 \end{equation}
where $\Lambda_4$ and $\Lambda_5$ are the 4 and 5 dimensional
cosmological constants and $\lambda$ is the tension of the brane.
The Friedmann equation is typically [2,19]

 \begin{equation}
 H^2=\frac{8\pi G}{3} \left (\rho +\frac{\rho^2}{2\lambda}\right
 )+\Lambda_4
 \end{equation}
 The tension
should be sufficiently large that the term quadratic in the energy
density should be suppressed by the time of nucleosynthesis [2];
and $\Lambda_4$ is taken to be negligibly small.

 The limitation that
result when using the FRW form for AdS means that any large
negative cosmological constant cannot be allowed to simply roam
free without some other mechanism counteracting the increased
gravitational force of attraction. Left alone this would simply
cause collapse to zero size.  The closed string modes having
matter  kinetic terms would diverge at such a point and cause a
resulting singularity. The brane-bulk model would fail  since all
the extra dimensions of M theory, or whatever have to be
adequately dealt with.

 The FRW AdS is also open so that the volume
is infinite, although one could take compact versions along the
lines of ref.[20]. However, the open geometry will generally cause
expansion while the negative cosmological constant will contract.
There is no stable equilibrium point as once a collapse ensues the
curvature cannot alone cause the Hubble parameter to change sign -
see eg. [3]. In order that the bulk be sufficiently long lived,
with lifetime $\sim |\Lambda_5|^{-1/2}$,the cosmological constant
must be exceedingly small even after allowing for the different
Planck scale in the five dimensional theory. As well as a fine
tuning that $\Lambda_5$ be sufficiently small,  this is
incompatible with requiring the volume $V$ to be only moderately
sized.

What is required is that the bulk be stabilized at around $\sim
mm$ scales by some other mechanism that presently seems unknown.
We have just pointed out the unjustified nature of imposing the
highly symmetrical AdS metric in such a scheme. This also goes
against the spirit of inflation, which is to try to deduce our
universe regardless of the  initial conditions, or at least with
only a reduced amount of symmetry - for a review see[21]. It is
also rather dubious to still invoke inflation on the brane, since
the initial underlying symmetry of the metric i.e. $SO(4,2)$ ,
does not strictly require, it at least for homogeneity. Using
inflation just to provide fluctuations seems extravagant since
setting up inflationary conditions generally means singularities
were earlier present [22]. Essentially because a cosmological
constant only dominates at larger scales. It is also not clear
that the justifications, for considering a displaced scalar field
as a source for chaotic inflation, are still valid. But, again
requiring the brane fields not to be in equilibrium is in stark
opposition to the statically imposed bulk. Although, matter is
supposedly constrained on the brane, this distinction cannot be
sustained as the Planck scale is approached and higher order
string corrections are required.

As an alternative one can try and produce an eternal brane by
achieving a bounce that prevents a singularity forming. For a
Reissner-Nordstrom AdS bulk the Friedmann equation becomes
modified, such that [19]
 \begin{equation}
 H^2+\frac{k}{a^2}=\frac{8\pi G}{3} \left (\rho +\frac{\rho^2}{2\lambda}\right )
 +\frac{M}{a^4}-\frac{Q^2}{a^6}
 \end{equation}

 where $M$ and $Q$ represent the mass and charge of the bulk
 space. In the Friedmann equation the $M$ term behaves like
 radiation while the charge $Q$ violates the weak-energy
condition. For a perfect fluid equation of state
$p=(\gamma-1)\rho$, a bounce can occur for matter softer that than
dust, i.e. $\gamma<1$.  For the closed case $k=1$ the resulting
perpetually oscillating model is like some simple version of the
Steady State universe - see eg. [23]. The advantage now is that
the total space-time is in  better harmony  with the PCP
principle. Since both sectors now, at least approximately, display
this principle. Other possible advantages of using such a charged
bulk have been considered in ref.[24], particularly that giving
$M$ and $Q$ a time dependence might help explain the required
value of $\Lambda_4$.  One problem still to overcome is that
perturbations will tend to grow rapidly during static or,
particularly, collapsing phases [25].

 We see the recent paper on Ekpyrotic universe [26] as another attempt  of
reconciling this dichotomy of the eternal nature of the bulk with
the apparent finite lifetime of our brane universe. The model is
very convoluted with ``our universe brane been sparked by
collision with a bulk brane that somehow left another hidden
brane''. Although we have grave doubts that this model is at all
natural, it also puts the total space-time in better harmony with
the PCP. Only the occasional emission of branes apparently
breaking this symmetry. Why this emission rate is not too slow or
too fast seems  one example of requiring  extra add-hoc
parameters.

 {\bf  Brane quantum cosmology}

 Can quantum cosmology be applied to justify this maximal
symmetry? A number of studies have considered Euclidean quantum
cosmology to be relevant to this problem and made certain
predictions [27], see also some related works [28]. This work
appears very muddled: should the bulk be assumed first and then
branes created, or should they both come into existence at the
same time?  In my view it only becomes justified to take the
Euclidean solutions when a forbidden region is present so that
quantum phenomena or tunneling could possibly be appropriate - see
[22] for more discussion of this more limited view. Others, take a
more liberal approach and consider Euclidean solutions even in the
absence of such a classical forbidden region. This seems
particularly the case of how quantum cosmology has been applied to
branes in the work [27]. We wish to broadly highlight a number of
problems that this seems to cause.

Recall in simple FRW models a forbidden region  is only present
when positive curvature and violation of the strong-energy
condition is present. Otherwise the initial singularity is not
isolated and simple classical evolution of the model can occur.
The archetypal example is the closed de Sitter model with
classical scale factor $a\sim \cosh t$, that might be formed by
quantum tunneling from the initial state of zero scale factor.
Since in the brane-bulk model there is no violation of the
strong-energy condition at least within the bulk, it seems, at
best,  contrived to invoke quantum cosmology. After all, since the
model displays the PCP with unchanging behaviour it presumably has
always existed. This would be like doing creation of  flat or
static de Sitter models. How is the value of the cosmological
constant determined ? The absence of a forbidden region also makes
it add-hoc whether Euclidean or Lorentzian times should be used
cf.[27]. This seems exacerbated when static metrics are also
present. Why was the 5 dimensional theory taken to be Euclidean at
some stage but now it must be Lorentzian: there seems no natural
anchor to fix the nature of the time parameter. Neither is the
curvature varying from a larger and more justifiably quantum
regime.

The related notion of signature change does not seem appropriate
either [29]. This was motivated by trying to avoid the need for
using Euclidean time
 when forbidden regions  occur at small scale factors [30].
 Although global AdS can have positive curvature it does not
 violate the strong-energy condition and any appropriate surface
 to change signature upon is not simply constrained to small size.
  Working with signature change in  FRW AdS type spaces  is even more fraught
   since they are curvature dominated at small scale factors.

 In another sense quantum cosmology
might still explain why this model exists compared to ones with
differing topologies but only if the universe was actually started
by tunneling from some other state and was free to tunnel where
the action, or whatever, was minimized. This avenue does not seem
particularly valid here since we expect the model to be the result
of evolution from some earlier higher dimensional string state. In
the usual applications of quantum cosmology, for example  in
scalar field type inflationary models, we assume the classical
general relativistic equations are roughly valid till the Planck
scale [14]. The unknown Planck region then becomes describable ,
or bypassed by ``instantons'',  by simply quantizing the usual
classical equations. The theory and resulting equations are never
assumed to be superseded with ``higher order'' corrections,
contrary to our present understanding of string theory with its
hierarchial ladder of descriptions.

Although the primeval string state could ultimately require
explanation from quantum cosmology, the current status of string
theory means this state is not yet even formulated. In the
meantime, we need to be  careful to  avoid using quantum cosmology
to supposedly predict conditions that  result from classical, or
actually string theoretic evolution from an earlier higher energy
state. Indeed the ultimate string theory would already include
quantum effects so further quantum cosmological reasoning would be
somewhat unwarranted. But because of  our present lack  of
understanding we are having to re-quantize the low-energy
classical string theories to pass back into the unknown regime. We
then of course hope they bear some relation to the real string or
M theoretic description in such a domain. The cardinal sin is to
move oppositely into the entirely or more classical
regime.\footnote{ Likewise it is rather perverse to consider
signature change in such a regime, when we have no reason  to
doubt the usual Lapse conventions [30].} There quantize and obtain
Euclidean ``instantons'', and apparently obtain that the universe
is initially created in such a state. There was once an argument
that claimed to predict a small $\Lambda$ for the present
universe, but implicitly assumed the universe had just now quantum
created itself into existence [31]. This type of error in quantum
cosmology seems easy to do, especially with static metrics that by
necessity are not being dominated by obvious quantum phenomena. A
similar example is that of obtaining the initial state for the
alternative string pre-big bang cosmology [32]. Here the starting
point is taken to be a near vacuum, classical state. Trying to
obtain such a state from quantum reasoning would simply appear
erroneous. Other principles such as ``string duality'' are instead
used to justify such a state. But these are not exact symmetries
and are now broken in our, also classical, universe [33]. This
reasoning also places quantum mechanics in a rather subsidiary
role for determining the initial state.

 We have
already remarked that quantum cosmology might only reasonably be
used to determine conditions on a suitable Cauchy
surface,\footnote{Or as remarked it cannot reasonably be expected
to ``cut and paste'' suitably expunged metrics together: this
being a highly severe form of horizon problem} which is
insufficient for the present case, with static AdS type metrics
required. Because we are using quantum cosmology to simulate this
unknown string evolution the requirement that a regulated
time-like singularity be present would seemingly put severe
constraint on the initial primeval string state. The cosmological
(brane-bulk) model could not be determined solely by causal
evolution from the primeval state. The relevant boundary
conditions would be being imposed continuously. This should be
contrasted with the usual inflationary models where the boundary
conditions are used solely to predict the initial classical
values. Whether their underlying principles such as `no boundary''
or ``only outgoing modes''[14,15] can be extended and applied to
non-globally hyperbolic spacetimes is an interesting point for
further study.

 We have only made some general remarks as to how the
issues should first be clarified before specific calculations can
be meaningfully attempted. But in summary it seems difficult to
conceive of how having both positive and negative cosmological
constants at the same time could result from some natural quantum
fluctuation. If the bulk was already present, the presence of a
large negative $\Lambda$ would supposedly suppress the creation of
a brane with a large positive $\Lambda$ compared to say, that from
an initially regular Minkowski space. Explaining the static bulk
itself does not seem an appropriate application  for quantum
cosmology.

{\bf Holography principle}

We also note that the static AdS metric plays a crucial role in
the holography principle of the AdS/CFT correspondence [34]. If
such metrics are not physically justifiable the correspondence
would only be of purely theoretical interest. The holography
principle $S\leq A$  would not of necessity be  valid in more
realistic cosmological models, for general reviews see [35]. It
has already been remarked that in the FRW AdS metric holography
principle is violated as collapse occurs [36]. This can be
understood since singularities take time to evolve and any $S>A$
can be postponed into creating a singularity in the far future of
the universe. We generally agree that the generalized second law
of thermodynamics is the more fundamental concept  for dynamic
models [37]. The holography principle only occurring for certain
restrictions, such as in static metrics or within the apparent
horizon size of FRW metrics where black holes can form ``quickly''
compared to the dynamical time scale of the background metric. We
suspect that these restrictions are built into the formulism of
the covariant holographic bound [38]. This takes holography as
primary, but it contains a rather convoluted recipe for
determining the relevant ``screens'' to apply the  principle to.

 {\bf Conclusions}

The brane world models with global AdS spacetimes are motivated by
trying to solve the hierarchy problem of particle physics: the
vast difference in the energy scales between particles and the
Planck scale. We should briefly mention that applying the notion
of ``varying constants''to brane models [39] is rather in conflict
with this basic aim. Increasing the speed of light makes the
Planck (mass) scale even larger  so exacerbating any existing
hierarchy problem [40].

 The brane  models typically use highly  symmetrical AdS
spaces and appear to
 explain our universe by symmetry breaking of the PCP to achieve our
 universe, some discussion of this possible paradigm shift can be found in ref. [41]
 . This approach treats supersymmetry as sacrosanct, but it
 becomes extremely contrived to obtain more usual cosmological
 models from  AdS, de Sitter and Minkowski or other static bulk metrics.
 It has earlier been pointed [42] out that it is highly restrictive
 that the bulk  possess such a high symmetry that the brane can display
 Lorentz symmetry $SO(3,1)$. We have rather emphasized how this is
 especially exacerbated when negative cosmological constants are present and
 the maximal symmetry
 is reduced to those like FRW. This also explains the difficulty
 of obtaining a  dynamical derivation [43]  of the ``see-saw''
 expression (4), as any negative $\Lambda$ will always dominate
 over matter to cause collapse of the bulk.

 The general drift in cosmology has rather been
  to reduce the initial imposed symmetry and
 instead utilize inflation as a means
  of securing homogeneity and isotropy to some sufficient degree.

Some partial resolution of the dilemma is to make both brane and
bulk eternal by allowing the brane to bounce instead of hitting
singularities. This, and the related Ekpyrotic model, at least
make the models more consistent. If the bulk is anyway going to
satisfy the PCP one can achieve almost this symmetry on the brane
by utilizing charged black holes in the bulk. The more general
bulk can spill over and effectively violate the weak-energy
condition within the brane. It would also be interesting to
consider the rotating AdS bulk case as this probably would put
dramatic constraints on the models. One major disadvantage of
accepting the PCP is that it would remain forever beyond the remit
of quantum notions to ever understand the reason for such a model.
Since its eternal existence means it is fundamentally a classical
construct.

On the other hand string theory is still very much developing.
Using maximally symmetric spaces with BPS states or Calabi Yau
style compactifications is probably too idealized and not yet
physically realistic. Just as full de Sitter isometry might not be
allowed in more realistic supergravity theories [44], we suspect
similar results will also be eventually found in static AdS,
especially if the cosmic censor is to be recovered in the low
energy limit, or an explanation produced for the ``cut and paste''
procedure.

  If brane type
models are to compete, from less structured or random initial
conditions, they somehow need to be formulated in more general
space-times. This makes the use of large negative cosmological
constants difficult, if not impossible, as they will generally
cause rapid gravitational collapse. Whether brane models with
stabilized large dimensions are possible remains an interesting
point of contention. It would generally be the case that
stabilizing extra dimension at small scales would be easier since
more quantum effects, like  the Casimir effect, can be invoked.
More usual quantum cosmology might then be applied to such schemes
- see e.g.[45], or something more exotic like quantum crumpling
might be appropriate [46] . Since there are presumably 7
dimensions to be adequately dealt with starting from 11
dimensional M theory; it might just be the case that they are all
curled up the same way. This unfortunately could simply be at too
small a scale for realistic experimental observation.

 {\bf Acknowledgement}\\ I should like to thank Daniel Chung,
 Roberto Emparan, Andrei Frolov, Christophe Grojean,
  Gary Horowitz, Robert Low, Roy
 Maartens and David Matravers  for helpful advice.
\newpage

{\bf References}\\
\begin{enumerate}
\item L. Randall and R. Sundrum, Phys. Rev. Lett. 83 (1999) p. 4690.
\\
L. Randall and R. Sundrum, Phys. Rev. Lett. 83 (1999) p.3370.
\item J. Cline, preprint hep-ph/0001285\\
D.A. Easson, preprint hep-th/0003086.
\item W. Rindler, ``Essential Relativity 2nd edn.'' (
Springer-Verlag: New York) 1977.
\item G.F.R. Ellis and T. Rothman, University of Cape Town report
1986 ( unpublished) \\ P. Anninos, R.A. Matzner and T. Rothman,
Phys. Rev. D 43 (1991) p. 3821.
\item M.P. Ryan and L.C. Shepley, `` Homogeneous Relativistic
Cosmologies'' (Princeton University Press : Princeton)1974.\\
 G.F.R. Ellis and D.R. Matravers, in `` A random walk'', eds.
N. Dadhich et al. (Wiley Eastern, New Delhi) 1985
\item B. Allen, Phys. Rev. D 32 91985) p.3136.
\item M.S. Madsen and P. Coles, Nucl. Phys. B 298 (1988) p. 2757.
\item S.W. Hawking and G.F.R. Ellis, ``The large scale
structure of space-time'',
Cambridge University press (1973).
\item G.T. Horowitz and D. Marolf, JHEP 9807 (1998) p.014. \\preprint hep-th/9805207.
\item M. Banados, M. Henneaux, C. Teitelboim and J. Zanelli, Phys.
Rev. D 48 (1993) p.1506.\\ M. Banados, C. Teitelboim and J.
Zanelli, Phys. Rev. D 49 (1994) p.975.
\item K. Koyama and J. Soda, Phys. Rev. D 62 (2000) p.23502.
\item Li-Xin Li, Phys. Rev. D 59 (1999) p.084016.
\item S.J. Avis, C.J. Isham and D. Storey, Phys. Rev. D 18 (1978)
p.3565.
\item J.J. Halliwell, in ``Quantum cosmology and baby universes''
eds. S. Coleman et al. (World Scientific: Singapore) 1991.\\ D.L.
Wiltshire, in  ``Cosmology the physics of the universe'', eds. B.
Robson et al. (World Scientific, Singapore) 1996. \\ also as
preprint gr-qc/0101003.
\item A. Vilenkin, preprint gr-qc/9812027.
\item R. Penrose, in ``General relativity'' eds. S.W. Hawking and
W. Israel ( Cambridge University Press: Cambridge) 1979.
\item J. Wudka, Phys. Rev. D 36 (1987) p.1036.
\item R. Emparan, JHEP 9906 (1999) p.036.
\item C. Barcelo and M. Visser, Phys. Lett. B 482 (2000) p. 183.\\
 preprint hep-th/0004056.
\item J.P. Luminet, Phys. Rep. 254 (1995) p.135.
\item D.S. Goldwirth and T. Piran, Phys. Rep. 214 (1992) p.223.
\item D.H. Coule, Phys. Rev. D 62 (2000) p.124010.
\item F. Hoyle, G. Burbidge and J.V. Narlikar, ``A different
approach to cosmology'' ( Cambridge University Press: Cambridge)
1999.
\item C. Csaki, J. Erlich and C. Grojean, preprint hep-th/0012143.
\item A.B. Batista, J.C. Fabris and S.V.B. Goncalves, Class.
Quant. Grav. 18 (2001) p.1389.
\item J. Khoury, B.A. Ovrut, P.J. Steinhardt and N. Turok,
preprint hep-th/0103239.
\item J. Garriga and M. Sasaki, Phys. Rev. D 62 (2000) p.043523.\\
S. Nojiri, S.D. Odintsov and S. Zerbini, Class. Quant. Grav. 17
(2000) p.4855.\\ A.S. Gorsky and K.G. Selivanov, preprints hep-th/
0005066 and  hep-th/0006044.\\ K. Koyama and J. Soda, Phys. Lett.
B 483 (2000) p.043501. \\ L. Anchordoqui, C. Nunez and K. Olsen,
JHEP 0010 (2000) p.050.\\
 S.W. Hawking, T. Hertog and H.S. Reall,
Phys. Rev D 63 (2001) p.083504.

\item E.I. Guendelman and A.B. Kaganovich, preprint gr-qc/9408031.\\
A. Davidson, preprint gr-qc/9710005\\ A. Davidson, D. Karasik and
Y. Lederer, Class. Quant. Grav. 16 (1999) p. 1349.\\ F. Darabi,
W.N. Sajko and P.S. Wesson, Class. Quant. Grav. 17 (2000)
p.4357.\\ P.I. Fomin and Yu.V. Shtanov, preprint hep-th/0008183.
\item M. Mars, J.M.M. Senovilla and R. Vera,
  preprint
gr-qc/0012099.
\item G.F.R. Ellis, A. Sumeruk, D.H. Coule and C. Hellaby, Class.
Quant. Grav. 9 (1992) p. 1535.\\
 A. Carlini, D.H. Coule and D.M. Solomons, Mod.
Phys. Lett. A 11 (1996) p.1453.
\item A. Strominger, Nucl. Phys. B 319 91989) p.722\\
D.H. Coule, Mod. Phys. Lett. A 10 (1995) p.1989.
\item M. Gasperini and G. Veneziano, Astro. Phys. 1 (1993) p.317.
\item D.H. Coule, Class. Quant. Grav. 15 (1998) p.2803.
\item J.M. Maldacena, Adv. Theor. Phys. 2 (1998) p.321.
\item L. Susskind, preprint hep-th/9409089.\\
 D. Bigatti and L. Susskind, hep-th/0002044.\\
L. Smolin, preprint hep-th/0003056.
\item N. Kaloper and A.D. Linde, Phys. Rev. D 60 (1999) p.4967.
\item R. Easther and D.A. Lowe, Phys. Rev. Lett. 82 (1999) p.4967.
\item R. Bousso, JHEP 9907 (1999) p.004.
\item E. Kiritsis, JHEP 9911 (1999)p.022.\\
 S.H.S. Alexander, preprint hep-th/9912037.
\item D.H. Coule, Mod. Phys. Lett. A (1999) p.2437.
\item T. Banks, preprint hep-th/9911067.
\item D.J.H. Chung, E.W. Kolb and A. Riotto, preprint
hep-ph/0008126.
\item T. Boehm, R. Durrer and C. van de Bruck, preprint
hep-th/0102144.
\item J. Maldacena and C. Nunez, preprint hep-th/0007018.
\item E. Carugno, M. Litterio, F. Occhionero and G. Pollifrone,
Phys. Rev. D 53 (1996) p. 6863.
\item R. Mansouri and F. Nasseri, Phys. Rev. D 60 (1999) p.123512.
\end{enumerate}
\end{document}